# Effects of using different plasmonic metals in metal/dielectric/metal subwavelength waveguides on guided dispersion characteristics

Ki Young Kim

Department of Physics, National Cheng Kung University, 1 University Road, Tainan 70101, Taiwan, Republic of China

**Abstract**

The fundamental guided dispersion characteristics of guided light in a subwavelength dielectric slit channel embedded by two different plasmonic metals are investigated when varying the gap width. As a result, an overall and salient picture of the guided dispersion characteristics is obtained over a wide spectrum range below and above the plasma frequencies of the two different plasmonic metals, which is important preliminary information for analyzing this type of subwavelength waveguide. In particular, the effects of using two different metals on the guided mode dispersions are emphasized in comparison with the effects of using the same plasmonic metal cladding.

## 1. Introduction

Recently, various types of waveguiding structure using the concept of surface plasmon polaritons (SPP) [1] have been intensively studied due to their subwavelength guiding ability of nanoscaled light [2-11], application to novel functioning photonic and optoelectronic devices [12-20], and enhanced optical transmission when using cross sections of these waveguides [20-23]. Among these subwavelength guiding structures, the fundamental guiding properties of metal/dielectric/metal (MDM) [or metal/insulator/metal (MIM)] waveguides have been rigorously studied [4-8] and also widely utilized in many applications [13-19] due to their structural simplicity as regards fabrication [13]. Moreover, MDM waveguides are able to support a quasi-transverse electromagnetic (TEM) mode [4], corresponding to the fundamental TEM mode of a parallel plate waveguide (PPW) in microwave technology [24], which is an important advantage for guiding subwavelength light, as it allows propagation, regardless of the gap width between the two metals.

Research on plasmonic MDM structures dates back almost 40 years, when Davis and Tamir theoretically analyzed dispersions in real plasma gaps [25] and Economou studied an insulating film between two semi-infinite metals [26]. Yet, there was no discussion on the possibility of subwavelength guidance. However, subsequent advancements in full-wave simulation, nanofabrication, and nanoscale measurement techniques have revealed more detailed properties of subwavelength MDM structures [5-7] and their novel photonic device applications [14-19]. While most recent studies have focused on MDM structures with the same metal cladding on both sides, a few studies have also investigated MDM structures with a combination of different metallic materials. For example, Prade *et al.* studied a dielectric slab waveguide with a different metal cladding on each side [27], and this hetero-metal concept was then used to propose novel metal heterowaveguides for the nanometric focusing of light [28, 29]. In addition, Shin *et al*. used hetero-MDM structures to break the symmetry of the field profiles in a perfect imager [30], and the same structures were also used to demonstrate negative refraction in visible frequencies [14]. Nonetheless, when considering such definitive demands on hetero-MDM structures in the fields of surface plasmonics or nanophotonics, the subwavelength guidance characteristics of a hetero-MDM structure need to be thoroughly investigated. Yet, available research on fundamental guided dispersions, an important guideline when designing or analyzing photonic guiding structures, is still insufficient. In particular, the gap-width dependence on the guided dispersions over a wide frequency spectrum is still incomplete and needs to be investigated further.

Accordingly, this paper explores the guided dispersion of subwavelength MDM waveguides when varying the gap width in order to obtain a more comprehensive understanding of the existing guided modes. The guided mode propagation is investigated, including the fast and slow wave characteristics over a wide spectrum range below and above the plasma frequencies of the two different metals. Plus, the effects of using a different metal cladding on each side are discussed in detail through a comparison of homo-MDM subwavelength waveguides [4] with the same gap widths. While this study uses silver and aluminum to obtain the guided mode characteristics, the present results could also be directly applied to the design and analysis of hetero-MDM subwavelength waveguides using other combinations of metals, and utilized as guidelines when designing subwavelength photonic devices adopting the same subwavelength hetero-MDM structure. It should be noted that the plasmonic material parameters for both metals are based on the lossless Drude model, and provide a clear picture of the guided dispersion characteristics of hetero-MDM subwavelength waveguides.

## 2. Subwavelength metallic slit structures and plasmonic material parameters

Figures 1 (a) and (b) show a cross-sectional schematic view with the coordinates of the hetero-MDM structure, and the plasmonic dispersions of the dielectric constants with the refractive indexes of the plasmonic metals under consideration in this paper, respectively. The light propagation is along the $z$-direction and the gap width between the two metals is $d$, as shown in figure 1(a). Aluminum (Al) and silver (Ag) were used for regions 2 and 3, respectively, and the dielectric constant of region 1 was assumed to be 1.0, that is, an air region. Thus, $\varepsilon_{r1}$, $\varepsilon_{r2}$, and $\varepsilon_{r3}$ represent the dielectric constants of air, Al, and Ag, respectively. The dispersive dielectric constant of two metals can be expressed as a simple lossless Drude model of $\varepsilon_{ri} = 1 - (\omega_{pi}/\omega)^2$, $(i = 2,3)$, as shown in figure 1(b), where the collision frequencies have been disregarded for simplicity. The bulk plasma frequencies for Al and Ag are $f_{p2} = \omega_{p2}/2\pi = 3570 \text{THz}$ and $f_{p3} = \omega_{p3}/2\pi = 2175 \text{THz}$, respectively [31]. Using these plasma frequencies in the lossless Drude model, the dielectric constants are then valid in the near- and far infrared regimes. In addition, Lorentzian resonance terms describing the interband transitions by electrons [32] need to be added to this simple Drude model to modify the optical behavior of real metals in frequency regimes like the visible and ultraviolet ranges [33]. However, while this modification of the simple Drude model facilitates a description of the optical properties of real metals for a broad spectrum range [5, 6], it can also diminish the more important effect of the dispersive plasmonic dispersion with a negative dielectric constant on the propagation characteristics. Nonetheless, other plasmonic structures using silver and with similar dimensions have already utilized this simple Drude model for quite a broad spectrum range and obtained significant physical insights associated with the plasmonic dispersion relation [34-36]. Thus, while recognizing the abovementioned sacrifice, it was decided to use the simple Drude model extending to the visible and ultraviolet spectrum ranges. The relative permeabilities for all regions were assumed to be 1.0, *i.e.*, $\mu_{r1} = \mu_{r2} = \mu_{r3} = 1.0$, and the conceptual transverse field profiles, designated by *A*, *B*, and *C* in figure 1 (a), are mentioned in the next section. Figure 1(b) shows the plasma resonance frequency for the plasmonic metals as $f_{r2} = f_{p2}/\sqrt{2} = 2524.37\text{THz}$ and $f_{r3} = f_{p3}/\sqrt{2} = 1537.96\text{THz}$, respectively, where the dielectric constants become negative unities, plus the dashed lines are the refractive indexes from the positive dielectric constants above the plasma frequencies as denoted by $n_i\,(i=2,3)$. Both the plasma resonance frequency and the refractive index play an important role in the guided dispersions, which will be discussed in the next section.

The characteristic equations for a hetero-MDM structure can be easily derived from the standard steps of boundary-value problems in similar ways to the characteristic equations for traditional three-layer asymmetric dielectric slab waveguides [37] as follows,

$$\left(\frac{\varepsilon_{r1}^2}{k_1^2} + \frac{\varepsilon_{r2}\varepsilon_{r3}}{k_2 k_3}\right)\tanh(k_1 d) + \frac{\varepsilon_{r1}}{k_1}\left(\frac{\varepsilon_{r2}}{k_2} + \frac{\varepsilon_{r3}}{k_3}\right) = 0 \quad \text{for the SPP mode} \qquad (1)$$

$$\left(\frac{\varepsilon_{r1}^2}{k_1^2} - \frac{\varepsilon_{r2}\varepsilon_{r3}}{k_2 k_3}\right)\tan(k_1 d) + \frac{\varepsilon_{r1}}{k_1}\left(\frac{\varepsilon_{r2}}{k_2} + \frac{\varepsilon_{r3}}{k_3}\right) = 0 \quad \text{for the PPW mode,} \qquad (2)$$

There are two distinctive transverse magnetic (TM) propagating modes: SPP and PPW modes, which propagate in slow-wave ($\beta > k_0$) and fast-wave ($\beta < k_0$) regions, respectively, where $\beta$ and $k_0$ are the propagation constants of the waveguide and free space (air), respectively. Plus, $k_1$, $k_2$, and $k_3$ are the transverse propagation constants in each region, given by $k_i = k_0 \left( \bar{\beta}^2 - \mu_{ri}\varepsilon_{ri} \right)^{1/2}$ ($i=1,2,3$), with the exception of $k_1 = k_0 \left( \mu_{r1}\varepsilon_{r1} - \bar{\beta}^2 \right)^{1/2}$ for the PPW mode, where $\bar{\beta}(=\beta/k_0)$ is the normalized propagation constant that is normalized by the free space wave number. The solutions of (1) and (2) are the normalized propagation constants with respect to their operating frequency providing the guided mode characteristics of the present hetero-MDM subwavelength waveguide, which will be discussed shortly.

## 3. Numerical results and discussion

Figure 2 shows the guided dispersion characteristics with several gap widths of 100, 60, 20, 10, and 5nm between the two different plasmonic metals. A 5nm gap width between the two metals was achievable with current fabrication technology [13]. A frequency range from 500 to 4000 THz was investigated, corresponding to 600 to 75nm in wavelength and covering parts of the visible to ultraviolet spectrum ranges. Two distinctive SPP mode solutions were obtained from (1): a $TM_0$ mode and $TM_1$ mode, corresponding to the transverse field profiles of *A* (quasi-symmetric, dominant even mode) and *B* (asymmetric, dominant odd mode) sketched in figure 1(a), respectively. Meanwhile, the first guided mode solution from (2), which was continuously connected to the previous $TM_1$ SPP mode solution at $\bar{\beta}=1.0$, belonged to a $TM_1$ PPW mode, corresponding to the sketched field profile of *C* in figure 1(a). The dispersion curves for the Al/Air/Ag waveguides lay between those for the Al/Air/Al and Ag/Air/Ag waveguides. Plus, the dashed regions are forbidden regions, where the guided modes cannot propagate, and the boundaries correspond to the dashed lines in figure 1(b).

It is well known that a $TM_0$ mode does not have a low-frequency cutoff. The present $TM_0$ mode existed completely within the slow wave region, *i.e.*, $\beta/k_0 > 1.0$, which is corresponding to the TEM mode of a PPW for microwave frequencies [4]. Also, the $TM_0$ modes were all forward waves, as the slopes of the dispersion curves were all positive. As the frequency increased (or the wavelength decreased), the normalized propagation constant ($\beta/k_0$) for the $TM_0$ mode also increased to the plasma resonance frequency of Ag, i.e., $f_{r3} = 1537.96 \text{THz}$, at which the normalized propagation constant had a very high value with a very low phase velocity ($k_0/\beta$). Meanwhile, the normalized propagation constant for the $TM_1$ mode, which extended across the slow and fast wave regions, had a very high value at the plasma resonance frequency of Al, i.e., $f_{r2} = 2524.37 \text{THz}$. Thus, despite the use of two different metals in the MDM structure, the $TM_0$ and $TM_1$ modes were tightly bound to the lower and higher plasma resonance frequencies, respectively, thereby differing from a homo-MDM structure [4] and surface plasmonic waveguide structure using the same metal with different dielectrics [38].

As the gap width between the two metals decreased from 100 to 5nm, the normalized propagation constant for the $TM_0$ mode maintained its tendency to increase with the operating frequency in the dispersion curves. Meanwhile, the low-frequency cutoff ($\beta/k_0 = 0$) for the $TM_1$ mode shifted toward a higher frequency regime until reaching the plasma frequency of Ag, i.e., $f_{p3} = 2175 \text{THz}$, above which the refractive index of Ag becomes positive. In the case of the homo-MDM structure with a narrow slit, the $TM_1$ mode solution existed in the form of a backward wave with a negative slope in the dispersion curve, between the plasma resonance frequency and the plasma frequency of the metal, i.e., $f_{ri} < f < f_{pi} (i = 2,3)$, where the dielectric constant was between negative unity to zero, as seen in figures 2(c) to (e) [dotted and dashed lines]. In contrast, in the case of the hetero-MDM structure, since the plasma resonance frequency of Al was greater than that of Ag, i.e., $f_{r3} > f_{p2}$, the solution had no choice but to be a forward wave with a positive slope for the dispersion curve, as shown in figures 2(c) to (e) [solid lines]. As a result, while the $TM_1$ modes of the Al/Air/Al and Ag/Air/Ag waveguides had mostly backward waves, the $TM_1$ modes of the Al/Air/Ag waveguides had partly forward waves, due to the use of two different metals in the subwavelength MDM structure. Furthermore, whereas the $TM_0$ modes could not exceed the plasma resonance frequency of Ag, *i.e.*, $f_{r3}$, the TM$_1$ mode was able to exceed the plasma resonance frequency of Al in the form of backward waves, as shown in figure 2(e), thereby enabling forward and backward waves to coexist with a very narrow gap width. Thereby, the zero group velocity at 2648.76 THz as well as at 2524.37 THz has been observed for the current lossless case. However, this might not be occurred when the absorptions in metals are introduced. The zero group velocities at 2524.37 THz in other gap dimensions are also expected to be moderated more or less resulting not being the zero group velocities anymore, which can also be found in a relevant plasmonic lossy slab structure [39]. More detailed study on how the guided dispersion characteristics of the present structure can be affected from two different amounts of metal losses has been left for further work. According to the results previously reported in [4], it could be expected that higher order TM modes and TE modes can only exist in the fast wave region and would be suppressed when the gap width is much smaller, although not shown here.

Figure 3 presents the evolutions of the TM$_0$ and TM$_1$ modes when varying the gap width, which is a repeat of figure 2, yet only for the hetero-MDM subwavelength waveguiding structures. Thus, figure 3 provides an overall and insightful picture of the guided mode dispersions of a hetero-MDM subwavelength waveguide, including the fact that the guided mode dispersions of the TM$_0$ and TM$_1$ modes are predominantly influenced by the metal with the lower and higher plasma resonance frequency, respectively, as mentioned previously. The TM$_0$ mode maintained its overall shape and exhibited an increasing normalized propagation constant with a decreasing gap width, whereas the dispersion curves of the TM$_1$ mode changed a lot more with the gap-width variations. In particular, the TM$_1$ mode was able to propagate in the form of a forward slow wave between the plasma frequency of Ag and the plasma resonance frequency of Al, where the dielectric constants of Ag and Al were $0 < \varepsilon_{r2} < 1$ and $-1 < \varepsilon_{r3} < 0$, respectively, which was not observed with the subwavelength homo-MDM structures.

Although this study did not treat propagation loss seriously, due to the use of the lossless Drude model for the plasmonic metals, the MDM structure is already known to be a highly lossy waveguide, especially when the phase velocity is very low, corresponding to a very narrow slit or the vicinity of the plasma resonance frequency. Yet, this situation may not be serious if the MDM structure is used as a transmitting medium for a very short distance or as part of a highly integrated nanoscale photonic device. Besides, using a gain medium as the center core of the MDM structure could somewhat mitigate this problem [40, 41], and its perspective is rather positive [42]. In this sense, current lossless guided dispersions can be of significance as a fundamental reference.

## 4. Conclusion

The fundamental guided dispersion characteristics of hetero-MDM subwavelength waveguides were investigated in detail. As a result, an overall and insightful picture of the guided mode dispersions was obtained over a wide frequency spectrum range, including below and above the plasma frequencies of the plasmonic metals, providing important preliminary information for a better understanding of hetero-MDM subwavelength structures. In contrast to a homo-MDM structure, it was found that the $TM_0$ and $TM_1$ modes were associated with the plasma resonance frequencies of the two different metals, and the forward slow wave of the $TM_1$ mode could even propagate in the frequency region where just one of the dielectric constants of the metals was positive. Thus, a hetero-MDM subwavelength waveguide can be used to break the field symmetry in a transverse direction, concentrate optical fields in a slow wave region, or suppress higher order modes in sophisticated photonic device designs, like bends, splitters, and multiplexers. Plus, in the case of subwavelength plasmonic waveguides with other geometries that use two different plasmonic metals, such as nanocoaxial waveguides, the different metal geometries with different plasmonic dispersions for the two different metals will also be expected to have a simultaneous effect on the guided dispersion characteristics. The results of this will be reported elsewhere.

**Acknowledgment**

This paper is dedicated to the memory of Mr. Beom-Seok Kim.

# Figure Captions

**Figure 1.** (a) Schematics of light propagation channel with different plasmonic cladding. Conceptual transverse field profiles for quasi-symmetric SPP, asymmetric SPP, and PPW modes are sketched in *A*, *B*, and *C*, respectively. (b) Plasmonic dispersion curves of Al and Ag with refractive index using simple Drude model. Solid and dashed lines are dielectric constants and refractive indexes of the plasmonic metals, respectively.

**Figure 2.** Guided dispersion characteristics with several gap widths between two different plasmonic metals.

**Figure 3.** Repeat of Fig. 2 for only subwavelength Al/Air /Ag waveguides with varying gap widths.

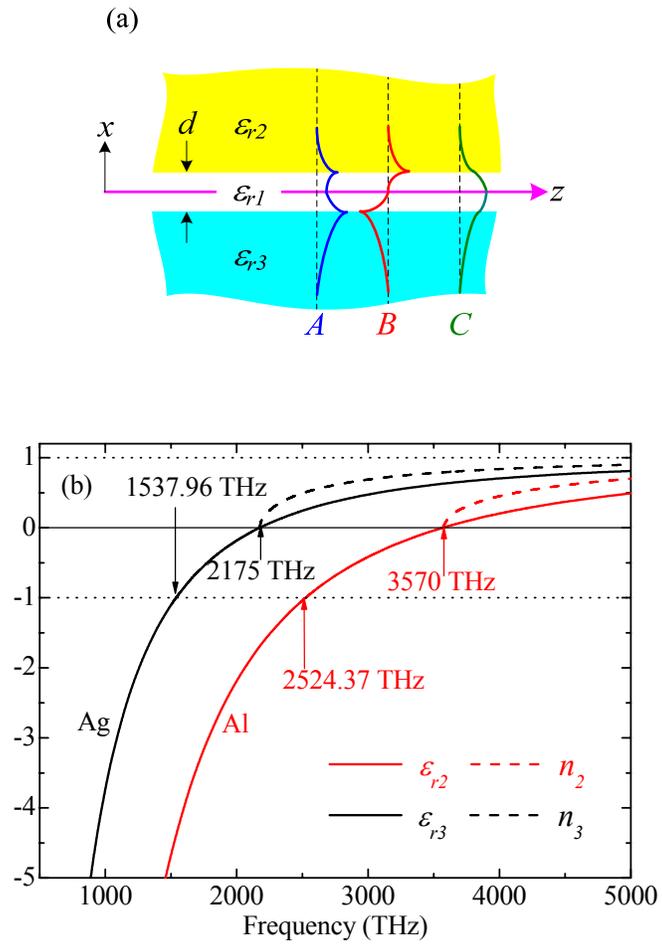

**Figure 2**

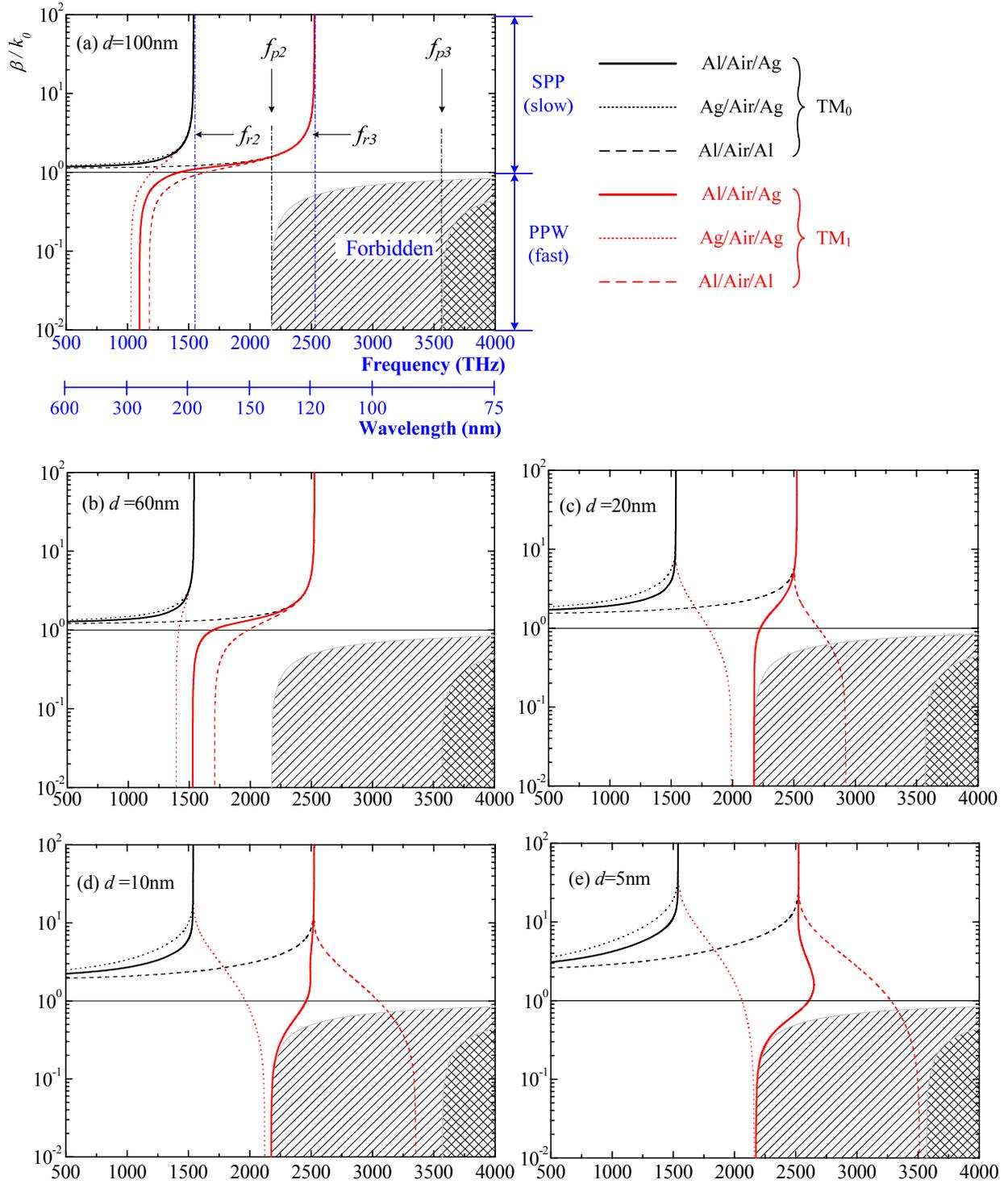

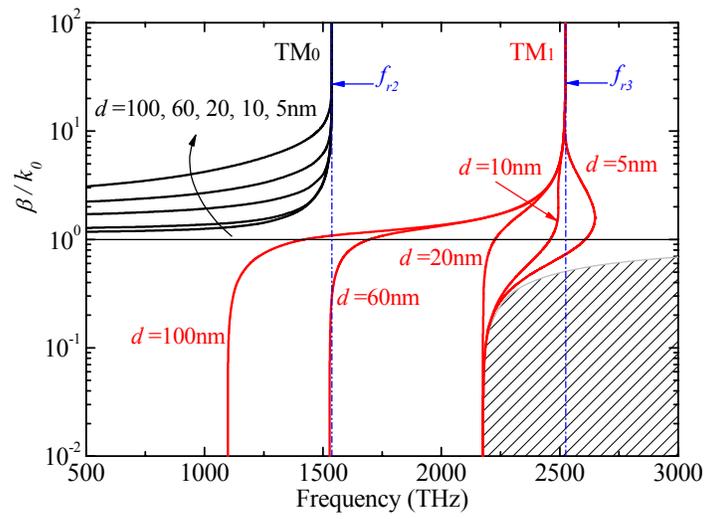